\documentclass[aps,prl,superscriptaddress,nofootinbib]{revtex4-2}

\usepackage{graphicx}
\usepackage{dcolumn}
\usepackage{bm}
\usepackage{multirow}
\usepackage{caption}
\usepackage{hyperref}
\usepackage{xcolor} 
\usepackage[symbol*]{footmisc} 
\begin{document}

\title{Mechanical control of the height distribution of adsorbed viral capsids}

\author{Yeraldinne Carrasco Salas\textsuperscript{*}}
\affiliation{CNRS, ENS de Lyon, LPENSL, UMR5672, 69342 Lyon cedex 07, France}
\author{Kassandra G\'erard\textsuperscript{*}}
\affiliation{CNRS, ENS de Lyon, LPENSL, UMR5672, 69342 Lyon cedex 07, France}
\author{Lauriane Lecoq}
\affiliation{Molecular Microbiology and Structural Biochemistry UMR 5086 CNRS/Universit\'e de Lyon, Lyon 69367, France}
\author{Anna Salvetti}
\affiliation{INSERM, D\'epartement de la science ouverte, Paris, France}
\author{Fabien Montel}
\author{Cendrine Faivre-Moskalenko}
\author{Martin Castelnovo}
\affiliation{CNRS, ENS de Lyon, LPENSL, UMR5672, 69342 Lyon cedex 07, France}


\begin{abstract}
The height of viral particles adsorbed on solid substrates is governed by the equilibrium between adhesion energy and capsid elasticity. While the resulting height distribution has been proposed as a non-invasive proxy for viral stiffness, the physical origin of its broadening is unknown. In this work, we combine Atomic Force Microscopy (AFM) topography measurements of Adeno-Associated Virus (AAV8) and Hepatitis B Virus (HBV) with a theoretical shell-deformation model to identify the determinants of height dispersion.
By modeling the viral shell as an elastic body under adhesive load, we evaluate the relative contributions of thermal fluctuations and mechanical heterogeneity to the observed height dispersion. We demonstrate that thermal noise is insufficient to explain the width of the distribution. Instead, the data support a model where the dispersion in height arises from the intrinsic variability of capsid stiffness. This variability is associated to the surface inhomogeneity of identical capsids. Our results validate that, when this inhomogeneity is accounted for, the height distribution of adsorbed particles provides a quantitative measure of viral mechanics without the need for individual nanoindentation.
\end{abstract}
\maketitle
\noindent\footnotetext{These authors contributed equally to this work.}

\section{Introduction}

The mechanical properties of viral capsids have been studied for over a decade as an indicator of viral stability \cite{ivanovska2004,roos2012}. Indeed, during the replication cycle of most viruses, the viral capsid, considered as a large supramolecular complex, undergoes major conformational changes, ranging from complete disassembly to self-assembly, following particular cues of physical, or biochemical origin.
Atomic Force Microscopy (AFM) has emerged as a method to study  the mechanical properties  due to its nanometric spatial resolution and its ability to perform controlled nanoindentation experiments on individual viral capsids \cite{ivanovska2004}. 
In typical AFM-based mechanical measurements, viral capsids are adsorbed onto flat substrates to immobilize them. A narrow AFM tip is first used to acquire the large-scale topography, and subsequently to probe the elastic response of the capsids by nanoindentation. This protocol requires a surface able to bind (or adsorb) the capsids. Obviously, the adhesion of viral capsids to the surface may induce a deformation, even before indentation is performed. This deformation results from the balance between viral capsid elastic and adhesion properties. This has been evidenced experimentally in different studies by modulating the elastic properties of viral capsids without changing the nature and strength of adhesion \cite{llauro2016}, or by changing completely the nature of the surface for the same virus \cite{zeng2017}. In this last study, it has been suggested that measurement of height distribution of a population of adsorbed viral particles could be used as a proxy to estimate quantitatively the stiffness of capsids. While this idea is appealing, a quantitative understanding of height distributions remains incomplete. In particular, it is not clear whether the observed dispersion in particle height arises from thermal fluctuations around an equilibrium configuration, from variations in adhesion strength, or from intrinsic heterogeneity in capsid mechanical properties. In the present work, we further explore a quantitative model that relates the height distribution of adsorbed viral capsids to both their elastic properties and their adhesion to the substrate. We demonstrate that dispersion in capsid stiffness plays a dominant role in shaping the height distribution. Indeed, it has been observed in several works that depending on the orientation of the adsorbed capsids, the force-indentation curves have different slope and therefore different stiffness \cite{carrasco2006,carrasco2008,menou2021}. 

This article is organized as follows. In the next section, we present the experimental method and analysis that allow to measure height distribution of viral particles adsorbed on flat surfaces using Atomic Force Microscopy (AFM).  We obtained data on adeno-associated virus serotype 8 (AAV8) and hepatitis B virus (HBV) capsids, for which the characterization of their mechanical properties have been previously obtained \cite{menou2021}. In particular, they have similar sizes, but different mechanical properties. AAV is a non-pathogen virus that is used as gene therapy vector \cite{salvetti1998}, while HBV is a major human pathogen \cite{lecoq2018}. Then, we present the new theoretical analysis of the height distribution. Finally, we analyze the data at the light of the model and we draw conclusions.

\section{Materials and methods}
\subsection{AFM height distribution measurements}
\subsubsection{Sample preparation}
AAV8 vector and Cp183 HBV capsids were prepared and purified as in Bernaud \emph{et al} \cite{bernaud2018}.  Viral capsid solution containing AAV8 vector or Cp183 HBV capsids was diluted to a final concentration close to $10^{12}$ viral particles $mL$, in Tris 10 $mM$ $pH=7.4$ buffer containing  1 $mM$ $NiCl_2$. Immediately after, 5 $\mu L$ of this solution was deposited onto freshly cleaved mica disks. After 20 minutes of incubation to favor adhesion, 30 $\mu L$ of imaging buffer containing 1 $mM$ $NiCl_2$ were added to the sample and another 70 $\mu L$ to the AFM liquid imaging cell in order to image the capsids in Peak Force imaging in liquid mode.
\subsubsection{AFM imaging}
Images were obtained using peak force mode in liquid on a nanoscope V multimode 8 AFM from Bruker. The cantilevers used were Bruker ScanAsyst-fluid plus and Bruker SNL-10(A) have a triangular geometry with a nominal radius of 2 $nm$ and a nominal stiffness of 0.7 and 0.35 $N/m$ respectively. The images were squares of 2 $\mu m$ side, with a resolution of 512x512 pixels and scanned at 2 lines/sec. The maximal force applied was 1000 $pN$ for a Peak Force Amplitude of 50 $nm$. The tilt and low noise frequency in the AFM images were removed using the NanoScope Analysis program by a first or second order flatten under height condition. The image analysis method is explained in the supplemental materials \cite{Sup}, and it uses a home-made Matlab code, first described in \cite{bernaud2018}.

\subsection{Model}
\subsubsection{Homogeneous stiffness} We consider in this section a theoretical model aimed at describing the adsorption properties of viral capsids onto a flat substrate. Our strategy is to relate the resulting particle height to the balance between elastic deformation of the capsid and adhesion energy arising from local contact between capsid subunits and the substrate. The elastic properties of viral capsids have been shown to be correctly described by the continuous theory of thin shell elasticity \cite{Landau1975}. In the case of a homogeneous spherical
thin shell of thickness $h$ and radius $R$, the equilibrium configuration of the shell in the absence or in the presence of external constraints results  from the
balance of in-plane stretching/compression and out-of-plane
bending. These features are quantified by the introduction of two modulus: the $2D$ stretching/compression modulus as $\kappa_s = Y h$
and the $2D$ bending modulus (also known as the flexural rigidity) as $\kappa_b=Yh^3/[12(1-\sigma^2 )]$, where $Y$ is the Young modulus, and $\sigma$ is the Poisson ratio of
the material.
A dimensionless number also known as the
F\"{o}ppl-von Karman number (FvK) $\gamma=\kappa_s R^2/\kappa_b$ measures the balance between compression and bending \cite{Landau1975,komura2005}. This number can be used for example in order to determine how "sharp" will be the shape of the closed shell: for low values, the equilibrium shape is almost spherical, while the shape is faceted at large FvK numbers \cite{lidmar2003}. Notice that we do not take into account  the gaussian bending modulus, as it is not expected to contribute significantly to the energetic balance,  assuming that the shell deformation induced by adsorption does not change the topology of the shell. This contrasts with the approach of reference \cite{zeng2017}. In this other work, it is assumed that the the flat part of the shell is connected to the undeformed shell through a critical region, the  "rim". This region represents a discontinuity in the shell configuration, and therefore they take into account for the contribution of Gaussian curvature (consistent with the change of topology of the shell) and also a rim energy associated to the discontinuity.

Under the influence of a local external mechanical constraint, the initial configuration is changed. Beyond a critical deformation threshold, a buckling instability eventually occurs, 
resulting in a modified trade-off between bending and stretching to accommodate deformation. This particular case is considered in the supplemental data \cite{Sup}, and we checked that the results obtained here with unbuckled configuration are quantitatively unchanged. In most AFM studies dedicated to measure stiffness of viral capsid, the indentation is limited to the unbuckled regime \cite{pogorelov}.
\begin{figure}
    \centering
\includegraphics[width=0.8\linewidth]{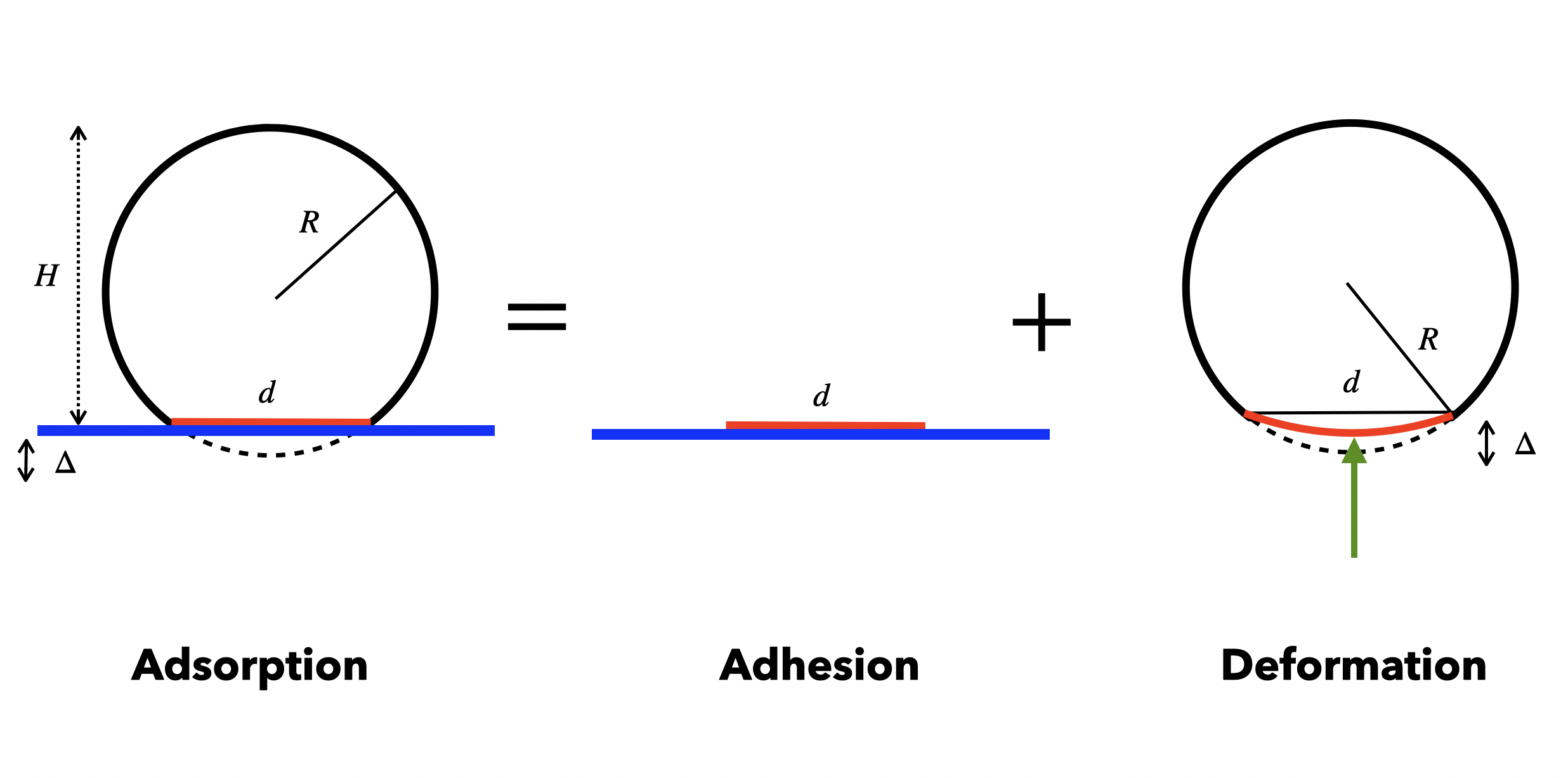}
    \caption{Geometry used in the energetic model in which particle adsorption is modeled as the superposition of deformation and adhesion. The red parts of the particles concentrate both the deformation and the interaction between proteins and the flat surface (in blue).}
    \label{Model geometry}
\end{figure}
As it was already mentioned, the estimation of the elastic cost can be obtained using a geometry as simple as the one used in standard calculation of indentation on a spherical shell, following the work of Komura \emph{et al.}\cite{komura2005}. The deformation is parametrized by the indentation length $\Delta$, and it spreads over the shell's surface on a region of area  $d^2\sim Rh$ \cite{Landau1975,lenz2008}. In this configuration, the shell is essentially the superposition of an undeformed spherical part and a flat part lying on the substrate (cf figure \ref{Model geometry}). 
The optimal elastic energy at fixed indentation length $\Delta$ is
\begin{eqnarray}
E_{elas} & \sim & \frac{Yh^2\Delta^2}{R} 
\end{eqnarray}
 This elastic behavior is valid as long as $\Delta<\Delta_*$, where the critical indentation length $\Delta_*$ being of the order of the thickness of the shell $\Delta_*\sim h$. Above this value for indentation, the configuration buckles. This configuration is described in the supplemental data \cite{Sup}. 
The variable of interest in our model is actually the height of the particle $H$, and its relation with indentation length is given by $H=2R-\Delta$. 

The adsorption cost is evaluated by adding an energetic gain to every subunit of the deformed area. This number is obtained by dividing the \emph{undeformed} area that is going to be deformed, by the typical area of the subunit $a^2$. By noting the energetic gain per subunit as $v$. The total energy is written in the unbuckled configuration as
\begin{equation}
    E_{tot}(H)  =  \frac{Yh^2(2R-H)^2}{R}-\frac{v R(2R-H)}{a^2}
\end{equation}
Notice that in the original work of Komura \emph{et al.}, the adsorption term is different in the unbuckled configuration \cite{komura2005}. This comes from the fact that their evaluation of the adsorption energy is based only on the deformed area and not on the number of subunits that effectively interact. Therefore, there is a tendency to include deformation as well in the interaction term. In their original calculation, the adsorption term does not depend on the value of the indentation $\Delta$. As a consequence, with Komura's choice, the total energy is minimal for $\Delta=0$, which is a counterintuitive result as it implies a shell with no deformation as the most favorable configuration.

Once the energetics of the adsorption has been modeled, it is possible to use this information in order to infer the height distribution. The simplest assumption is to suppose that the adsorption takes place under equilibrium conditions, \emph{i.e.} the particle explores all possible configurations (or height values) using thermal energy alone. Under this assumption, the height distribution obeys a Boltzmann distribution:
\begin{equation}
\label{Boltzmann}    P(H)\sim e^{-E_{tot}(H)/(k_BT)}
\end{equation}
where  $k_B$ is the Boltzmann constant, and $T$ the  temperature.
\begin{figure}
    \centering
\includegraphics[width=0.8\linewidth]{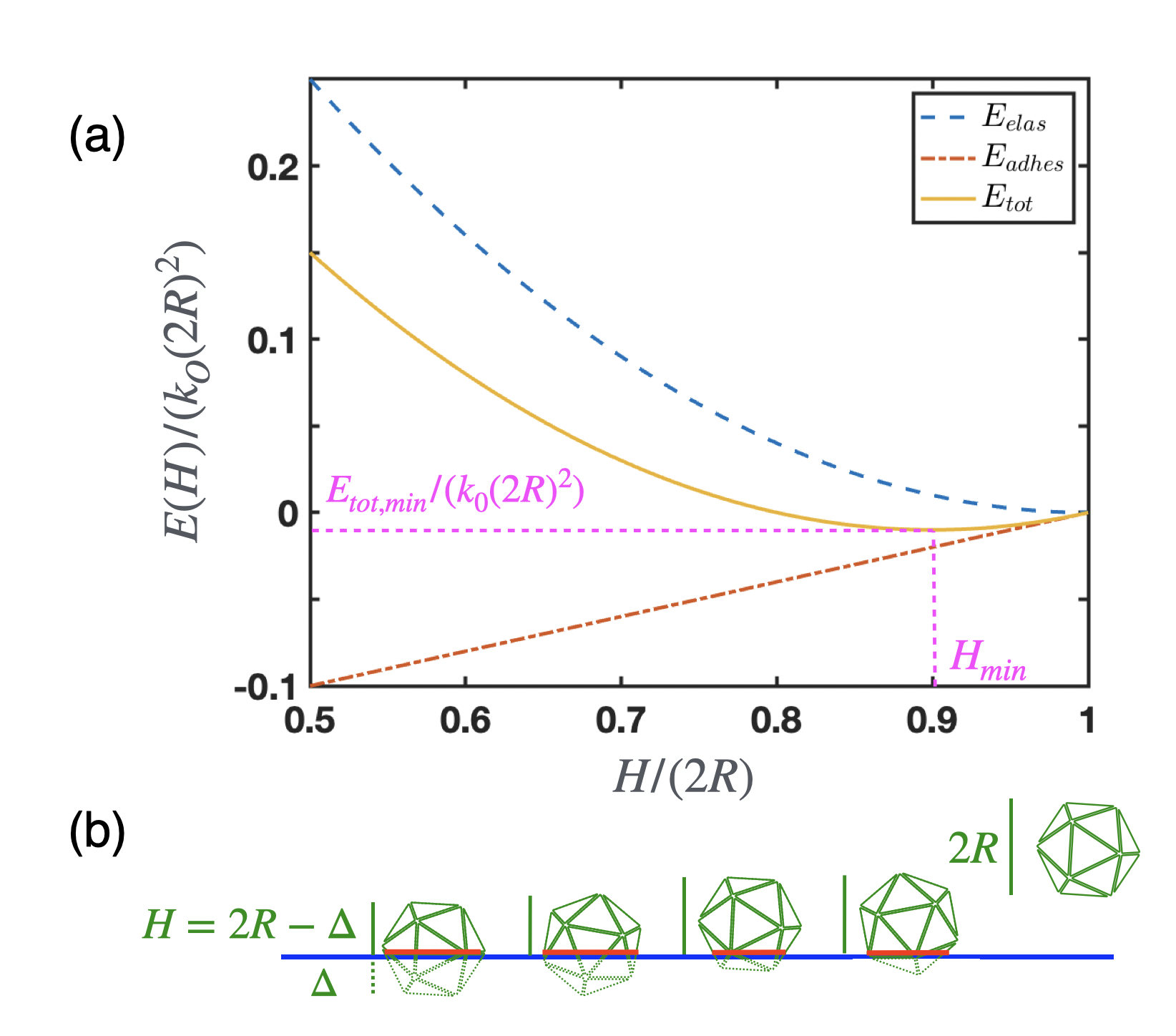}
    \caption{Energetic model of height distribution. (a) Total dimensionless energy (yellow continuous line) as function of relative height $H/(2R)$, together with its elastic (blue dashed line) and adhesive (red dot-dashed line) contributions. The chosen parameters are $\frac{v}{2a^2 k_0}=0.2$ and $k_0(2R)^2=k_BT$. (b) Diagram illustrating the origin of height distribution.}
    \label{Model diagram}
\end{figure}

In order to be able to compare the predictions of the model to experimental data, we found it more simple to rely on the estimation of the first two statistical moments of the height distribution, rather than trying to fit the distribution itself. The comparison of direct least-square fitting and using the method of moments is shown in the supplemental data \cite{Sup}. In order to do so, we approximate the height distributions using a saddle-point method. When the shell is not buckled, this is a trivial result as the energy is already quadratic, but it is useful when considering the non quadratic energy of buckled configuration (see supplemental data \cite{Sup}). This approach allows to obtain analytical expressions for the first two moments of the distribution. The idea of this approach is to approximate the total energy by a harmonic potential around its true minimum position:
\begin{equation}
E_{tot}(H)=E_{tot,min}+\frac{1}{2}\left(\frac{\partial^2 E_{tot}}{\partial H^2}\right)_{H=H_{min}}\left(H-H_{min}\right)^2    
\end{equation}
As a consequence, the first two moments of the height distribution Eq.\ref{Boltzmann} will be given by
\begin{eqnarray}
    \mu & = &  H_{min} \nonumber\\
    \sigma^2 & = & k_BT/\left(\frac{\partial^2 E_{tot}}{\partial H^2}\right)_{H=H_{min}}
\end{eqnarray}
In the case of the unbuckled configuration, the position and energetic value of the minimum are in this case given by
\begin{eqnarray}
   H_{min} & = & 2R-\frac{R^2v}{2h^2a^2Y}\\
    E_{tot,min} & = & -\frac{R^3v^2}{4h^2a^4Y}\\
    \left(\frac{\partial^2 E_{tot}}{\partial H^2}\right)_{H=H_{min}} & = & \frac{Yh^2}{R}
\end{eqnarray}

We notice that 
\begin{equation}
    \left(\frac{\partial^2 E_{tot}}{\partial H^2}\right)_{H=H_{min}}\sim \frac{E_{tot,min}}{(H_{min})^2}
\end{equation}
so that there are only two relevant parameters in order to characterize the distributions, namely the net energetic gain for adsorption and the optimal particle height. We provided a rigorous derivation of this result in the supplemental material \cite{Sup}. The typical energy is illustrated in  figure \ref{Model diagram}a.

Direct estimation of material properties like Young modulus $Y$ or shell thickness $h$ is not an easy task from an experimental point of view. Rather, AFM nano-indentation experiments allows to estimate typically the spring constant, or the stiffness,  associated to the whole particle, which contains informations on both of these parameters. Most nanoindentation measurements are performed in the linear regime of deformation, where the configuration is expected to be  unbuckled. In this case, the stiffness scales as \cite{Landau1975,menou2021}
\begin{equation}
\label{Stiffness}    k\sim\frac{Yh^2}{R}
\end{equation}
up to a numerical prefactor. Using this stiffness as representative of the elastic properties of the shell, rather than its Young Modulus, the first two moments of the height distributions in the unbuckled configuration are rewritten as
\begin{eqnarray}
\label{mu unb}    \mu & =  &  2R-\frac{R v}{2a^2k}\\
\label{sigma unb}    \sigma^2 & = & \frac{k_BT}{k}
\end{eqnarray}

\subsubsection{Inhomogeneous stiffness}
It is possible within our model to take into account for the dispersion in measured stiffness. Such a dispersion has been quantified recently\cite{menou2021}. It is related to the inhomogeneous local environment on the virus surface, which shows icosahedral symmetries. As particles are adsorbed to the surface with different orientations, the elastic response to adsorption is modulated, and this impacts the height distribution.  
Taking this feature into account, the observed population of capsids can be modeled as a mixture of subpopulations of particles with fixed stiffness $k_i$. As a consequence, the height distribution is now written as 
\begin{equation}
\label{mixture}    P_{mix}(H)=\sum_{i}p(k_i) P(H|k_i)
\end{equation}
where $p(k_i)$ is the weight of  viral capsids with stiffness $k_{i}$ among the total population, and each subpopulation contributes to the height distribution with the Boltzmann weight $P(H|k_i)$ previously described. The first moments of the total distribution are given by
\begin{eqnarray}
    \mu & = & \sum_i p(k_i)\mu_i \\
    \sigma^2+\mu^2 & = & \sum_i p(k_i)(\sigma_i^2+\mu_i^2)
\end{eqnarray}
where $\{\mu_i,\sigma_i\}$ are the mean and standard deviation of height for viral capsids with stiffness $k_i$. 
These formula can be made explicit in the case of unbuckled configuration:
\begin{eqnarray}
    \mu & = & 2R-\frac{Rv}{2a^2}\left\langle \frac{1}{k}\right\rangle\\
    \sigma^2 & = & k_BT\left\langle \frac{1}{k}\right\rangle+\frac{R^2v^2}{(2a^2)^2}\left(\left\langle \frac{1}{k^2}\right\rangle-\left(\left\langle \frac{1}{k}\right\rangle\right)^2\right) 
\end{eqnarray}
 with the following notation \begin{eqnarray}
\label{eq18} \left\langle \frac{1}{k}\right\rangle & = & \sum_i \frac{p(k_i)}{k_i}\\
\label{eq19} \left\langle \frac{1}{k^2}\right\rangle & = & \sum_i \frac{p(k_i)}{k_i^2}
\end{eqnarray}
In order to obtain the estimation of stiffness dispersion influence on height distribution,  we further assume that stiffness are normally distributed with a mean stiffness $k_0$ and standard deviation of stiffness $\gamma_k$, all other parameters being equal (radius, thickness). Then the first moments of height distribution can be  approximated as to the lowest order in relative stiffness dispersion $\alpha=\frac{\gamma_k}{k_0}$
\begin{eqnarray}
    \mu & = & 2R-\frac{Rv}{2a^2k_0}\left(1+\alpha^2\right)\\
    \sigma^2 & = & \frac{k_BT}{k_0}\left(1+\alpha^2\left(1+\left(\frac{Rv}{2a^2}\right)^2\left(\frac{1}{k_0\,k_BT}\right)\right)\right)
\end{eqnarray}
These relations can be combined in a way to remove the explicit dependence on adhesion parameter. To the lowest order, we obtain the following relation
\begin{equation}
\label{main prediction sigma}\sigma^2  =  \frac{k_BT}{k_0}(1+\alpha^2)+\alpha^2(2R-\mu)^2
\end{equation}
The first term represents the direct  contribution of mean stiffness to height dispersion, while the second term represents the influence of stiffness dispersion.


\section{Results and discussion}

\begin{figure}
    \centering
\includegraphics[width=0.8\linewidth]{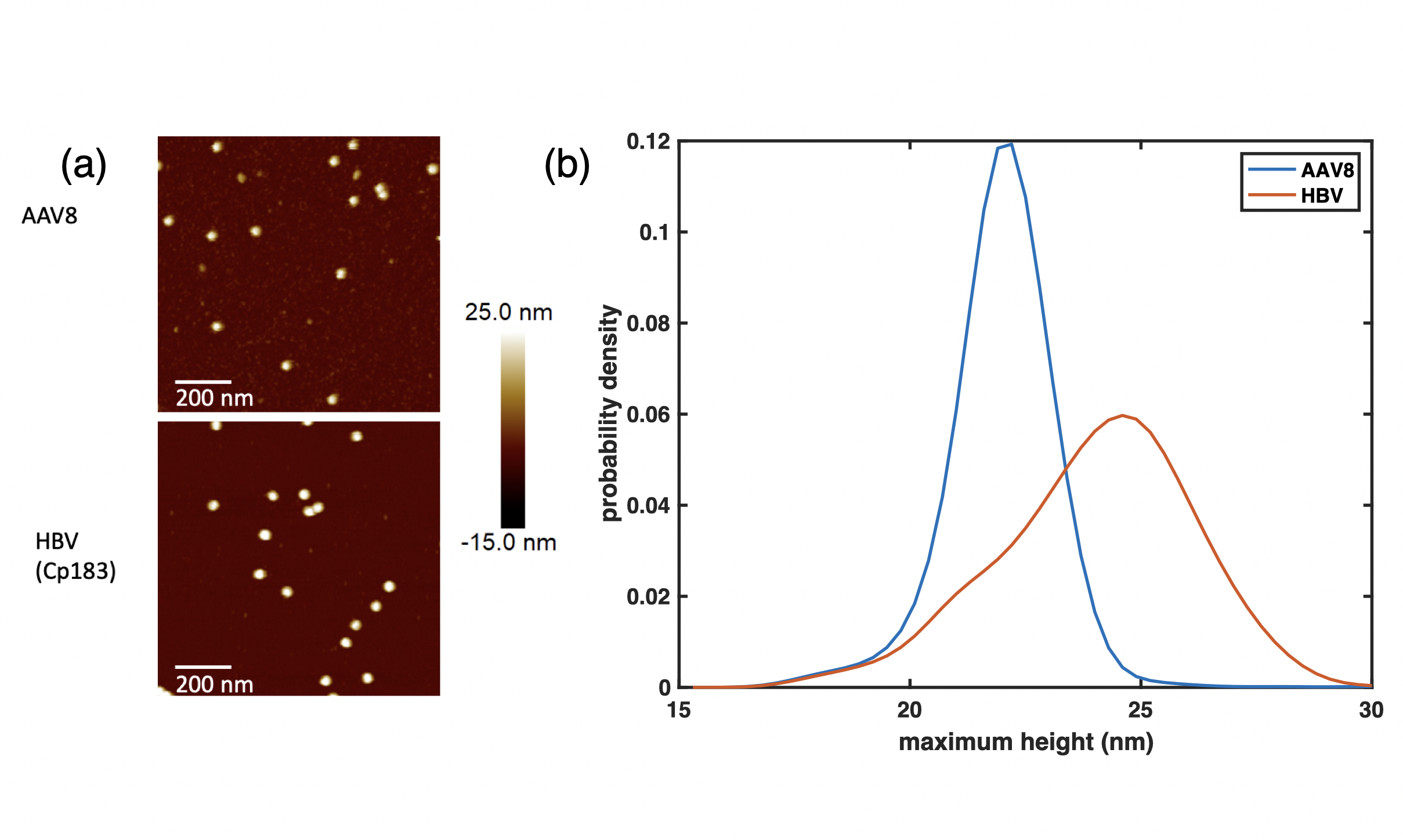}
    \caption{Images and height distributions for AAV8 and HBV. \emph{(a)} Images of AAV8  and HBV.  \emph{(b)} Height distributions for $N_{AAV8}=1026$ and $N_{HBV}=961$ capsids. }
    \label{Histogram data}
\end{figure}

The height distribution has been quantified for two viral capsids derived from AAV8 and HBV. This distribution is represented in figure \ref{Histogram data}. Our modeling strategy is to analyze the first statistical moments of these distributions in order to discuss which physical mechanisms, thermal fluctuations, adhesion or stiffness heterogeneity, can account for the observed dispersion, and to relate to the elastic parameters associated with the two viruses. Note that these distributions are asymmetric, with a small but non negligible proportion of events with low heights. 
These events are likely associated with strongly deformed capsids, which are not described within our linear elasticity model. Therefore our analysis will focus mainly on the first two statistical moments of the distributions, which are relevant in the framework of linear elasticity: the mean value and the standard deviation. Using the proposed model for these statistical moments of height distribution, we will address two questions: estimating what adhesion energy is necessary to reproduce the observed mean height once the stiffness is measured, and estimating the height dispersion using again the stiffness measurement. 
\subsection{Estimating adhesion energy}
First, it is possible to estimate the adhesion strength required in order to obtain the observed height distributions for the two types of viruses investigated.
 These two viruses have different nominal radii: $R_{AAV8}=12.5nm$ and $R_{HBV}=15nm$. The protein thickness of the shell are roughly identical $h_{AAV8}\simeq h_{HBV}\simeq 3nm$ \cite{bozic2013,wynne1999}. 
Estimation of the adhesion strength is made by using the height mean value and stiffness in the following equation
\begin{equation}
  \bar{v}=   2k_{0}\frac{2R-\mu}{2R}= 2 k_{0}\bar{\Delta}
\end{equation}
with $\bar{v}=\frac{v}{2a^2}$. 
The first statistical moments of these distributions are extracted in the table \ref{Table 1}, the relative deformation being defined as $\bar{\Delta}=\frac{2R-\mu}{2R}$.

\begin{table}[h!]
\begin{center}
\label{Table 1}
\begin{tabular}{| c |c | c | c | c | c |}
\hline
\, & & & & & \\
 \, & $k_0 (N/m)$ & $\gamma_k(N/m)$ & $\mu$ (nm)& $\sigma$ (nm) & $\bar{\Delta}_i$ \\
\hline
$AAV8$ & 0.8 & 0.35 & 22.0 & 1.2 & 0.12 \\ 
\hline
$HBV$ & 0.3 & 0.11 & 24.1 & 2.1 & 0.20 \\  
\hline
\end{tabular}
\end{center}
\caption{\label{Table 1} Experimental results for capsid stiffness $k$, the first two moments of the height distributions $\mu$ and $\sigma$ and the relative deformation.}
\end{table}

Using the measured mean value of stiffness, we find respectively $\bar{v}=0.19 J.m^{-2}$ $\simeq 48 k_B T/nm^{-2}$ and $\bar{v}=0.12 J.m^{-2}\simeq 30 k_B T/nm^{-2}$ for AAV8 and HBV. These adhesion parameters are slightly different, but of the same order of magnitude. This is not so surprising as it depends on the type of surface, the environmental conditions but also on the chemical nature of the subunits. The value of adhesion parameter is quite high for a molecular complex. Yet it is associated with the deformation of the capsid whose elastic properties are quite tight as well. Indeed, the conversion of mean stiffness from $J.nm^{-2}$ to  $k_BT.nm^{-2}$ gives $k_{AAV8}=0.8N.m^{-1}=200k_BT.nm^{-2}$ and $k_{HBV}=0.3N.m^{-1}=75k_BT.nm^{-2}$.
This means that the strong stiffness of viruses requires strong adhesion. It might be reasonable at this step to compare the value of adhesion energy found in the present work to the one obtained in the work of reference \cite{zeng2017}. However we postpone this discussion to the next subsection of \emph{Inhomogeneous stiffness} for reasons that will be become clear soon.
\subsection{Estimating height dispersion}
\subsubsection{Homogeneous stiffness} The consequence of combined strong adhesion and stiffness is that the width of height distribution is expected to be rather thin, as for both viruses the depth of energetic well of adsorbed configuration is of order $k_0 (2R-\mu)^2\sim 10^3 k_BT$.
Similarly, using Eq.\ref{sigma unb} for the  standard deviation of height, one obtains $\sigma\simeq 0.07nm$ for AAV8 and $\sigma \simeq 0.12nm$ for HBV, based on the measurement of the mean stiffness. These \emph{predicted} values are at least two orders of magnitude smaller than measured one. A partial conclusion at this step is that the observed height dispersion is not due to thermal fluctuations around the mean value. 
\subsubsection{Inhomogeneous stiffness} 
Using Eq.\ref{main prediction sigma}, it is possible to test the influence of stiffness dispersion on the height dispersion. The previous estimation of height dispersion corresponds to the $\alpha=0$ case (no stiffness dispersion). When the stiffness dispersion is not negligible, it is observed in Eq.\ref{main prediction sigma} that the height dispersion should increase linearly with the stiffness dispersion, with the following asymptotic law 
\begin{equation}
\label{Final prediction}\frac{\sigma}{(2R-\mu)}=\alpha=\frac{\gamma_k}{k_0}
\end{equation}
In other words, when the stiffness dispersion is taken into account, we expect the \emph{relative} height dispersion to equal the relative stiffness dispersion. This particularly simple relation is obtained asymptotically under the assumption of normal gaussian distribution of stiffness. Real measurements like the one performed in reference \cite{menou2021} have exhibited that stiffness distribution might be multimodal, reflecting different orientation for adsorbed viral particles. Nevertheless, in order to test the simple relation proposed in our work, we used the statistical parameters of the main stiffness peak.   Our measurements of relative stiffness dispersion for AAV8 gives $k_{AAV8}=(0.8\pm0.35)N/m$ and $\alpha_{AAV8}=0.44$. These data would correspond to height dispersion $\sigma_{AAV8,esti.}\simeq 1.3nm$ which is close the measured value $\sigma_{AAV,meas.}\simeq 1.2nm$. Similarly, for HBV capsids, we obtained $k_{HBV}=(0.3\pm0.11)N/m$ and  $\alpha_{HBV}=0.37$. These data would correspond to height dispersion $\sigma_{HBV,esti.}\simeq 2.2nm$, which is close the measured value $\sigma_{HBV,meas.}\simeq 2.1nm$. We checked that taking into account the multimodal feature of true stiffness distribution does not alter these estimations, as multimodality shifts both mean stiffness and its dispersion to higher values, but their ratio is roughly constant. Having established the correlation between experimental height dispersion and inhomogeneous stiffness, we can now discuss the value of adhesion energy found in the present work and compare them to the one obtained in reference \cite{zeng2017}. We obtained values of adhesion energy per area of 30-50 $k_BT/nm^2$, while the previous cited work obtained values of order 0.1 $k_BT/nm^2$.  In our case, these values are obtained using only the mean height, while in the other case the values have been obtained by fitting the whole height distribution, and therefore these values contain information on height dispersion as well. Therefore the discrepancy is not surprising: the "large value" of height dispersion is associated to small adhesion energy within the energetic model with homogeneous stiffness. Our approach introduces a new degree of freedom, \emph{i.e.} the stiffness dispersion, to account for the height dispersion and therefore the adhesion energy is different.
\subsubsection{Influence of polygonal shape} 
At this point, one might argue that capsid shells are not purely spherical, and that the polygonal shape might contribute as well to the measured height dispersion. We can test quantitatively this scenario by calculating first the typical height dispersion assuming that each capsid has the shape of an undeformed truncated icosahedron (fullerene-like particles like C-60 molecule, or a soccer ball with flat faces). The difference of height between configurations with facing pentamers (\emph{i.e.} the capsid lays on a pentamer) and facing hexamers (\emph{i.e.} the capsid lays on a hexamer) is given exactly by 
\begin{equation}
\Delta H=H_{penta}-H_{hexa}=\frac{2R}{(1+9 \phi^2)^{1/2}}\left(\frac{(42+41\phi)^{1/2}}{5^{1/2}}-3^{1/2}\phi^2\right)\simeq0.02 (2R)
\end{equation}
with $\phi$ the Golden Ratio, and $R$ the circumradius of sphere passing through all vertices of the truncated icosahedron \cite{Fullerene}. For both viruses, this distance would correspond to a height dispersion of order $0.3-0.4nm$. This value is much smaller than the measurement, and therefore the shape of the capsid cannot explain the observed height dispersion.

A second sanity check on the effect of capsid shape can be performed along the following argument. It was already mentioned that the stiffness of a thin shell is related to its radius by the relation Eq.\ref{Stiffness}. If we consider a variation in the effective radius (due to the polygonal shape) we can compute the expected stiffness variation, and compare it the measured stiffness dispersion. The equation \ref{Stiffness} implies the following relationship
\begin{equation}
    \frac{\delta k}{k}=\bigg |\frac{\delta R}{R} \bigg |\simeq 0.02
\end{equation}
This value has to be compared to the value of parameter $\alpha$ introduced earlier, which is about 0.3-0.4, \emph{i.e.} one order of magnitude larger. Therefore the polygonal shape does not contribute significantly to the stiffness dispersion, and therefore to the height dispersion. Finally, we tested in the supplemental material whether the height dispersion could be related directly to viral particle polydispersity \cite{Sup}.

\subsubsection{Comparison to other data}
The prediction of Eq.\ref{Final prediction} can also be tested on another set of data published in the literature. As an example, Zeng \textit{et al.} measured two height distributions of adsorbed Brome Mosaic Virus (BMV) on two different surfaces \cite{zeng2017}: HOPG and Mica/Mg(II). Since the virus is identical for the two height distributions, the relative stiffness dispersion should also be the same. As a consequence, Eq.\ref{Final prediction} predicts that the height dispersion should increase as the mean height decreases. This is indeed qualitatively confirmed by their data. 

\section{Conclusions}
In the present work, we have proposed a quantitative model for height distribution of adsorbed viral particles on a flat substrate, and its correlation with stiffness distributions. Our model takes into account the energetic balance between elasticity and adhesion. The main elastic parameter of the model is the stiffness of the capsid, which is easily measured using AFM nanoindentation. 

There are two main conclusions for our study. First, the capsids' mean height is related both to their stiffness and their adhesion strength with respect to the surface, and these contributions cannot be separated.  
This contrasts with the classical analysis of indentation experiments. Indeed, indentation is performed onto adsorbed particles, in order to avoid any motion of the particle during adsorption. Yet, for a small deformation, the capsid sticks to the regime of linear elasticity, and the slope of force-deformation curves is representative of its stiffness, without any contribution from the adsorption process. Therefore this is the standard and objective way of measuring stiffness. But when no external force is imposed on the particle, the adhesion and stiffness are already coupled at a linear level of the modeling and their contribution cannot be disentangled, just like in our analysis.

\begin{figure}
    \centering
\includegraphics[width=0.8\linewidth]{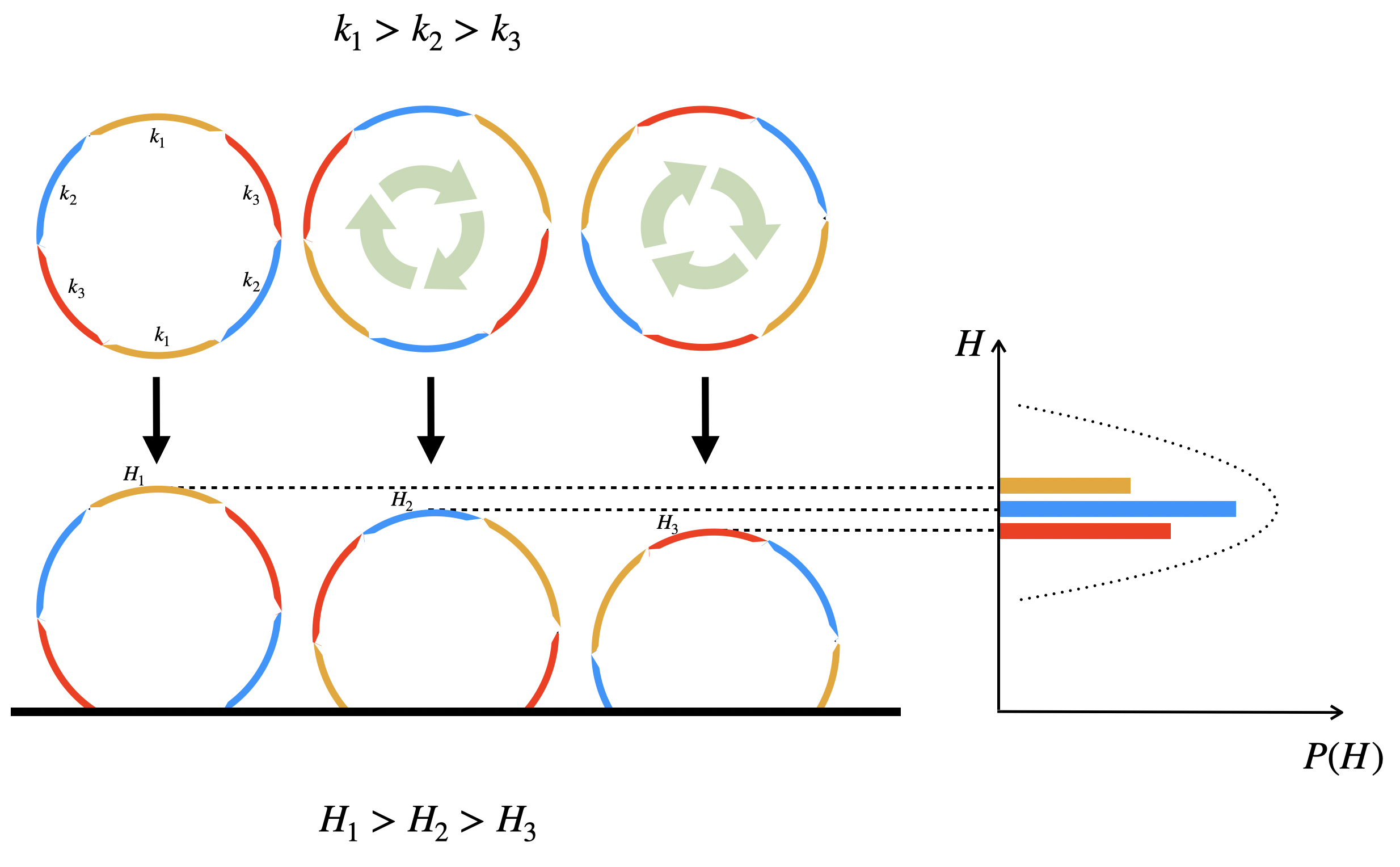}
    \caption{Illustration of the proposed mechanical control mechanism. For illustrative purposes, the particle to be adsorbed has three distinct regions of stiffness (represented by the three colors: yellow, blue, and red). During adsorption,  an identical adsorption condition results in different adsorption orientations and  therefore induced deformations. Consequently different heights are expected. This yields a typical height histogram, shown on the right.}
    \label{interpretation}
\end{figure}

Second, the only information that can be extracted from height distribution, which is independent of adsorption properties, is its relative standard deviation (the ratio between standard deviation and mean value). Indeed, we have shown both theoretically and experimentally that this measure is strongly correlated with the relative standard deviation of stiffness distribution, and that it is not related to the strength of adsorption. The qualitative picture is as follows: although all capsids are expected to be identical, their sensitivity to deformation is not uniform across their entire surface. Consequently, depending on the orientation of the capsid during adsorption, it will be adsorbed to a greater or lesser extent, resulting in a broadening of the height distribution that correlates with the stiffness distribution (Figure \ref{interpretation}).

As a conclusion, we found that the large dispersion of height distributions of viral capsids adsorbed on a substrate is not dominated by the energetic balance between elasticity and adhesion, but rather by the dispersion of stiffness measured. And this dispersion traces back to inhomogeneous surface properties of viral capsids. Our results indicate that height dispersion contains quantitative information about mechanical heterogeneity, and can therefore be used as an indirect probe of stiffness dispersion, complementary to nanoindentation measurements. Although developed here for viral capsids, the proposed framework could be applicable to other soft nanoscale objects adsorbed on substrates, for which mechanical heterogeneity may similarly affect adsorption-induced deformations. In this case, the knowledge of this mechanical heterogeneity can be used in Eq. \ref{eq18}-\ref{eq19} to relate stiffness dispersion and height dispersion.

\bibliographystyle{apsrev4-2}  
\bibliography{virus}

\end{document}


\title{Mechanical control of the height distribution of adsorbed viral capsids - Supplemental materials}

\author{Yeraldinne Carrasco Salas\textsuperscript{*}}
\affiliation{CNRS, ENS de Lyon, LPENSL, UMR5672, 69342 Lyon cedex 07, France}
\author{Kassandra G\'erard\textsuperscript{*}}
\affiliation{CNRS, ENS de Lyon, LPENSL, UMR5672, 69342 Lyon cedex 07, France}
\author{Lauriane Lecoq}
\affiliation{Molecular Microbiology and Structural Biochemistry UMR 5086 CNRS/Universit\'e de Lyon, Lyon 69367, France}
\author{Anna Salvetti}
\affiliation{INSERM, D\'epartement de la science ouverte, Paris, France}
\author{Fabien Montel}
\author{Cendrine Faivre-Moskalenko}
\author{Martin Castelnovo}
\affiliation{CNRS, ENS de Lyon, LPENSL, UMR5672, 69342 Lyon cedex 07, France}

\maketitle
\noindent\footnotetext{These authors contributed equally to this work.}

\section{Supplemental Data}
\subsection{AFM image analysis}
\begin{figure}
    \centering
\includegraphics[width=0.8\linewidth]{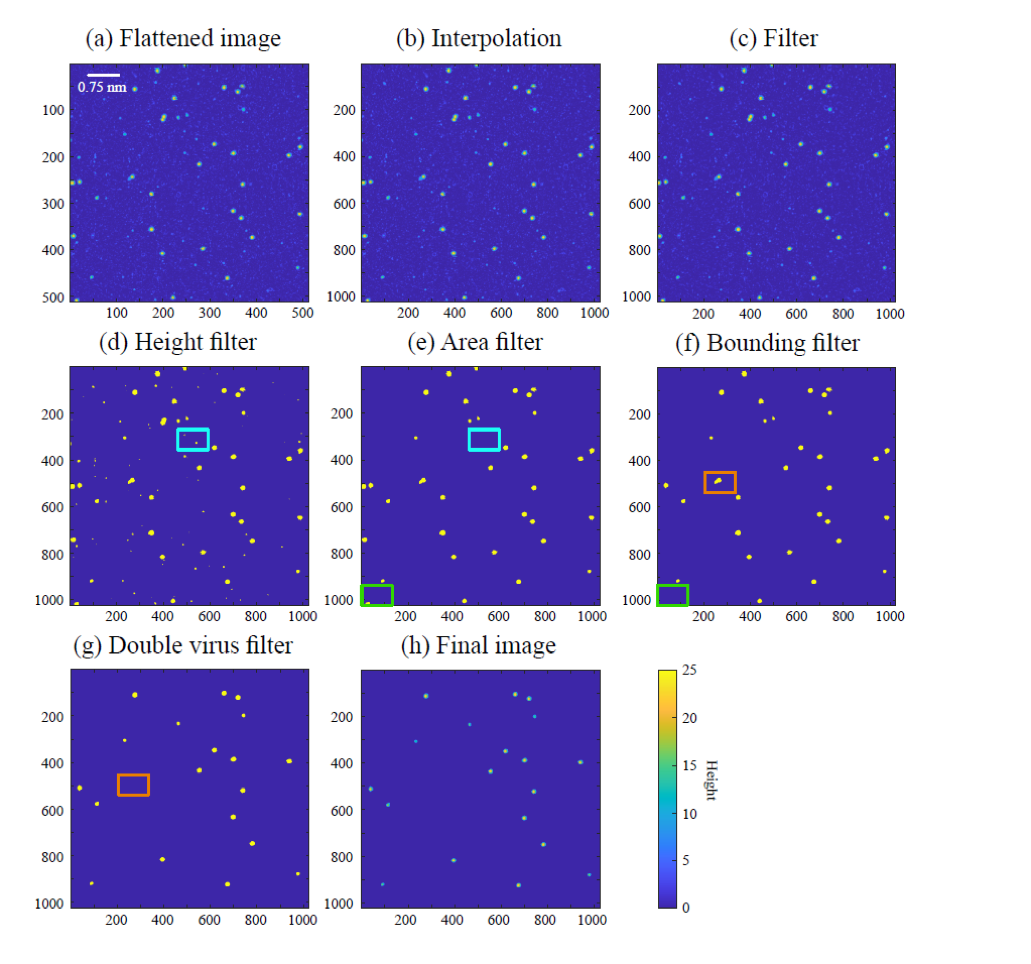}
    \caption{Image analysis steps: (a) Flattened image coming from NanoScope Analysis program (b) Interpolation and (c) noise filter performs an image. The height filter (d) and area filter (e) remove all the small detections associated to noise and/or impurities of the image. (f) Bounding filter removes capsids touching the edge of the image. (g) Double virus filter removes overlapped capsids. (h) Final processed image for capsid morphological parameter analysis. Here, all non-zero value pixels represent the analyzed capsids. The colored boxes represent the comparison between each step effect of the analysis with the previous image.}
    \label{image analysis}
\end{figure}

To obtain the capsid morphological properties, we applied a similar image processing analysis to that published by Bernaud \emph{et al} \cite{bernaud2018}. Briefly, the interpolation process and the spatial filters applied were as follows: the height threshold used was 5 $nm$, which allowed us to separate the capsid surface (Fig. \ref{image analysis}d). Then, an area filter was applied, with minimum and maximum area thresholds of 500 and 2000 $nm^2$, respectively (Fig. \ref{image analysis}e). In the ideal case of no tip dilatation effect on the capsid, the area of the capsid 2D projection is approximately 650 $nm^2$ (for an AAV8 capsid radius of $R_{AAV8} = 12.5 nm$). Therefore, the minimum area of a capsid should be around this value. The filters used until now do not allow us to select isolated capsids as some capsids are still in contact with each other.
To eliminate dimers or oligomers of capsids, we used a parameter called the fractal number, defined as $F=P^2/(4A)$, where $P$ and $A$ are the object perimeter and area, respectively. Ill-defined capsids for $F <0.8$ and agglomerated capsids for $F >1.2$ were removed (Fig. SUP \ref{image analysis}f-g). Capsids touching the edge of the AFM image were also removed. Then, we applied a capsid filter, \emph{i.e.} we selected structures with a height greater than 10 $nm$ this ensures that only intact capsids are selected. The colored boxes represent the effect of each step of the analysis on the previous image. Finally, we multiplied the treated, binarised image by the original image to create an image with the original height value for each pixel belonging to well-defined, isolated capsids (Fig. SUP \ref{image analysis}h). Then we calculated the different morphological properties, such as the height, area and volume, of each capsid. To obtain the capsid height, we identified the maximum $2*2$ pixels area value. AFM height measurment resolution is on the order of 0.2 $nm$ in our imaging conditions.

\subsection{Buckling model}
\begin{figure}
    \centering
\includegraphics[width=0.8\linewidth]{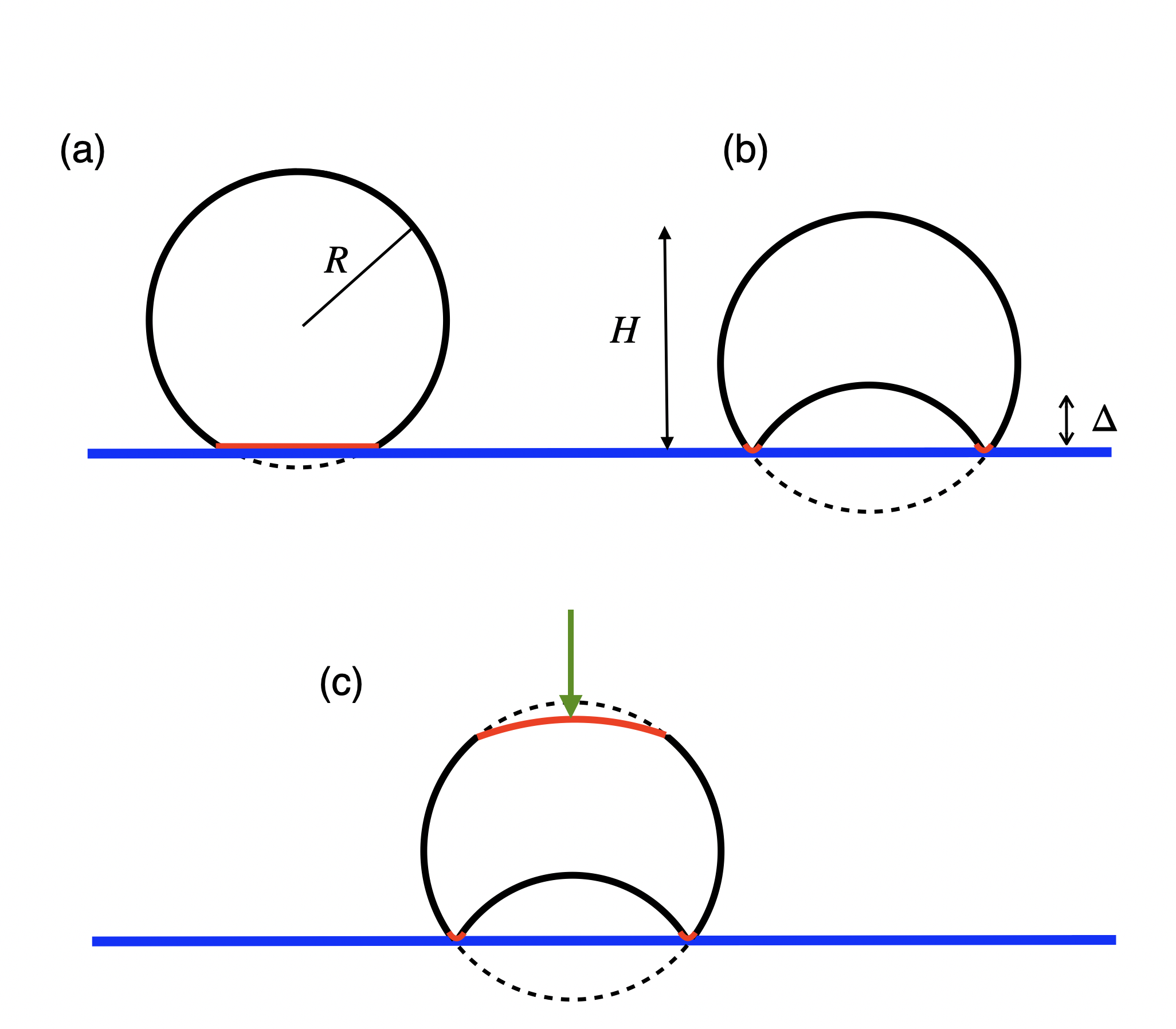}
    \caption{Illustration of the geometry proposed within our model. (a) \emph{Unbuckled} configuration, (b) \emph{buckled} configuration, (c) mixed configuration in the presence of both adsorption and indentation. The deformed area is shown  in red. The dashed line shows the original undeformed configuration. The area of this lower part is given by $A=2\pi R \Delta$.}
    \label{schema}
\end{figure}

Under the influence of a local external mechanical constraint, the initial configuration is changed, and beyond a critical deformation threshold, a buckling instability occurs, 
resulting in a modified trade-off between bending and stretching to accommodate deformation. As an example, the application of a local force on the top of a spherical shell will first induce a localized continuous deformation, which is dominated by increasing both in-plane compression and bending, and then it will buckle inward beyond a critical force/deformation. This is illustrated in figure SUP\ref{schema}a and SUP\ref{schema}b, when the deformation is caused by adsorption for example. This final configuration allows to focus the mechanical stress onto the rim connecting the inverted cap to the original shell while relaxing the bending energy in the inwardly buckled part, and therefore it lowers the overall stress. This result has been quantified in several theoretical works, and we will use the derived scaling law for elastic energy in the optimal configuration at fixed indentation length \cite{Landau1975,pogorelov,lenz2008,menou2021}.
The deformation only extends to a \emph{finite} area (for moderate deformations), such that elastic balance can be applied to this finite area, the remaining area of the particle being undeformed. At this point, we stress out that within an indentation experiment, the top and the bottom of particle are not necessarily in the same configuration: indeed, the bottom part of the particle is in contact with the subtrate, and its deformation is due to the adsorption process, while the top part is deformed due to the external load. If the adsorption energy is large enough, it is possible to have a bottom buckled configuration, and a top unbuckled configuration (cf figure SUP \ref{schema}c).
 In the post-buckling configuration,  the shell is again composed of an undeformed spherical part, and the deformed part is now an inward spherical buckled part (cf figure SUP \ref{schema}c). 
The optimal elastic energy at fixed indentation length $\Delta$ is
\begin{eqnarray}
E_{elas} & \sim & \frac{Yh^2\Delta^2}{R}  \,\,\,\,\,\,\,\,\,\,\mathrm{if}\, \Delta<\Delta_*\nonumber\\
 & \sim & \frac{Yh^{5/2}\Delta^{3/2}}{R}  \,\,\,\mathrm{if}\, \Delta>\Delta_*\nonumber
\end{eqnarray}
where the critical indentation length $\Delta_*$ being of the order of the thickness of the shell $\Delta_*\sim h$. 
The adsorption cost is evaluated by adding an energetic gain to every subunit of the deformed area. This number is obtained by dividing the \emph{undeformed} area that is going to be deformed, by the typical size of the subunit $a$. By noting the energetic gain per subunit as $v$. The total energy is written in the unbuckled configuration as
\begin{equation}
    E_{tot}^{(u)}(H)  =  \frac{Yh^2(2R-H)^2}{R}-\frac{v R(2R-H)}{a^2}
\end{equation}
 and in the buckled configuration as
\begin{equation}
    E_{tot}^{(b)}(H)  =  \frac{Yh^{5/2}(2R-H)^{3/2}}{R}-\frac{v R(2R-H)^{1/2}h^{1/2}}{a^2}
\end{equation}

In order to be able to compare the predictions of the model to experimental data, we found it more accurate to rely on the estimation of the first two statistical moments of the height distribution, rather than trying to fit the distribution itself. In order to do so, we approximate the height distributions using a saddle-point method. This allows to obtain analytical expressions for the first two moments of the distribution. The idea of this approach is to approximate the total energy by a harmonic potential around its true minimum position:
\begin{equation}
E_{tot}^{(i)}(H)=E_{tot,min}^{(i)}+\frac{1}{2}\left(\frac{\partial^2 E_{tot}^{(i)}}{\partial H^2}\right)_{H=H_{min}^{(i)}}\left(H-H_{min}^{(i)}\right)^2    
\end{equation}
As a consequence, the first two moments of the height distribution  will be given by
\begin{eqnarray}
    \mu_{(i)} & = &  H_{min}^{(i)} \nonumber\\
    \sigma_{(i)}^2 & = & k_BT/\left(\frac{\partial^2 E_{tot}^{(i)}}{\partial H^2}\right)_{H=H_{min}^{(i)}}
\end{eqnarray}
In the case of the unbuckled configuration, the position and energetic value of the minimum are in this case given by
\begin{eqnarray}
   H_{min}^{(u)} & = & 2R-\frac{R^2v}{2h^2a^2Y}\\
    E_{tot,min}^{(u)} & = & -\frac{R^3v^2}{4h^2a^4Y}\\
    \left(\frac{\partial^2 E_{tot}^{(u)}}{\partial H^2}\right)_{H=H_{min}^{(u)}} & = & \frac{Yh^2}{R}
\end{eqnarray}
while in the case of buckled configuration, these reads like:
\begin{eqnarray}
    H_{min}^{(b)} & = & 2R-\frac{R^2v}{3h^2a^2Y}\\
    E_{tot,min}^{(b)} & = & -\frac{2}{3^{3/2}}\frac{R^2v^{3/2}}{h^{1/2}a^3Y^{1/2}}\\
    \left(\frac{\partial^2 E_{tot}^{(b)}}{\partial H^2}\right)_{H=H_{min}^{(b)}} & = & \frac{3^{3/2}}{2}\frac{aY^{3/2}h^{7/2}}{R^2 v^{1/2}}
\end{eqnarray}
In both cases, we notice that 
\begin{equation}
    \left(\frac{\partial^2 E_{tot}^{(i)}}{\partial H^2}\right)_{H=H_{min}^{(i)}}\sim \frac{E_{tot,min}^{(i)}}{(H_{min}^{(i)})^2}
\end{equation}
so that there are only two relevant parameters in order to characterize the distributions, namely the net energetic gain for adsorption and the optimal particle height.

Direct estimation of material properties like Young modulus $Y$ or shell thickness $h$ is not an easy task from an experimental point of view. Rather, AFM nano-indentation experiments allows to estimate typically the spring constant, or the stiffness,  associated to the whole particle, which contains informations on both of these parameters. Most nanoindentation measurements are performed in the linear regime of deformation, where the configuration is expected to be  unbuckled. In this case, the stiffness scales as \cite{Landau1975,menou2021}
\begin{equation}
k\sim\frac{Yh^2}{R}
\end{equation}
up to a numerical prefactor. Using this stiffness as representative of the elastic properties of the shell, rather than its Young Modulus, the first two moments of the height distributions in the buckled configuration are rewritten as
\begin{eqnarray}
\label{mu u}    \mu_{(u)} & =  &  2R-\frac{R v}{2a^2k}\\
\label{sigma u}    \sigma_{(u)}^2 & = & \frac{k_BT}{k}
\end{eqnarray}
for the unbuckled configuration, and 
\begin{eqnarray}
\label{mu b}    \mu_{(b)} & =  &  2R-\frac{R v}{3a^2k}\\
\label{sigma b}    \sigma_{(b)}^2 & = & \frac{2}{3^{3/2}}\frac{k_BT R^{1/2} v^{1/2}}{ak^{3/2}h^{1/2}}
\end{eqnarray}
in the buckled configuration.

\subsection{Fitting the distribution}
\begin{figure}
    \centering
\includegraphics[width=0.8\linewidth]{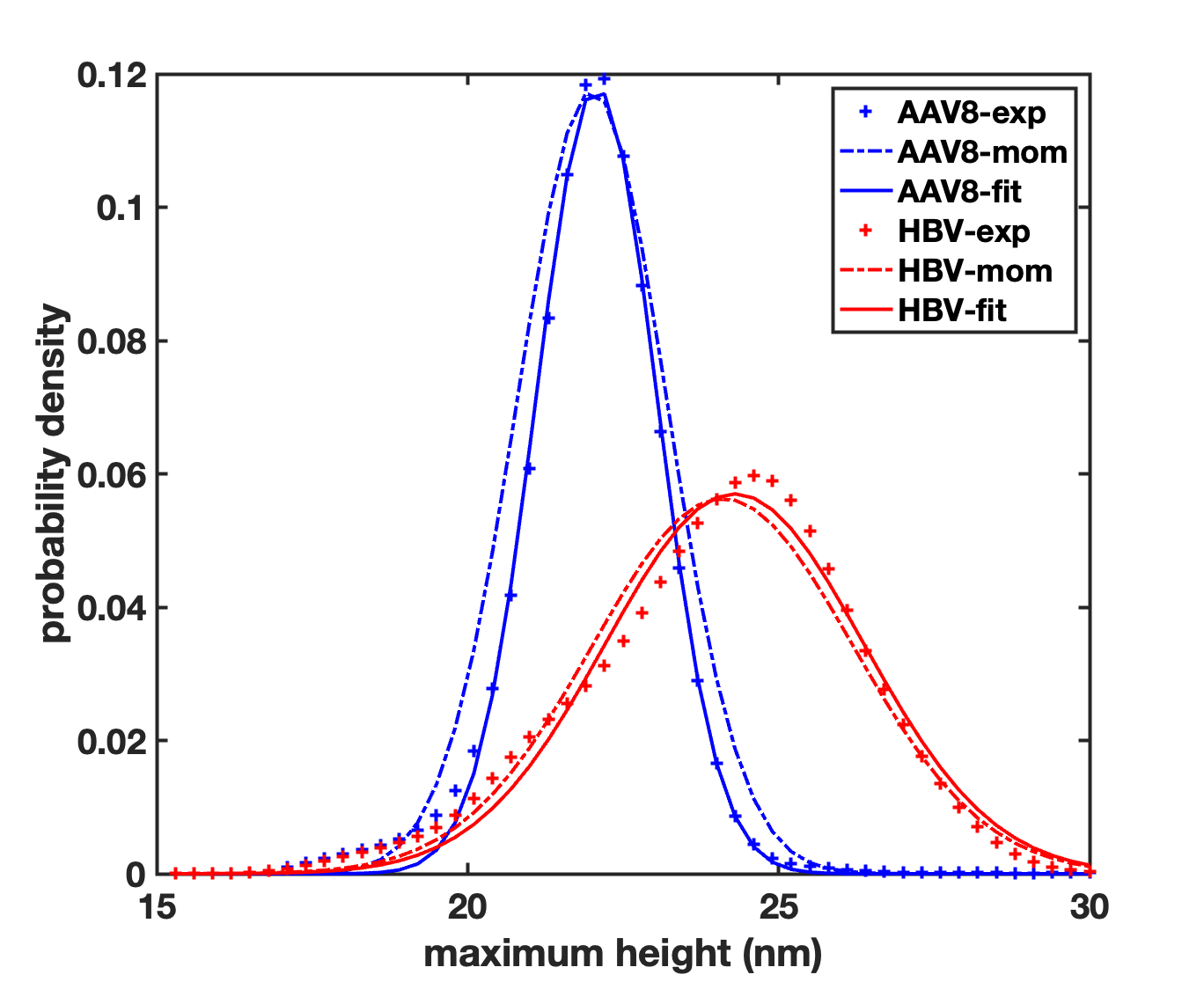}
    \caption{Height distribution for AAV8 \emph{(blue)} and HBV \emph{}(red)}
    \label{fit}
\end{figure}

In this section, we present the comparison of two procedure in order to extract the model parameters from the experimental distribution: the method of moments, which has been used in the main text, and the direct least-square fitting of the distribution. The value of the extracted parameters are shown in the following table. In both case, parameters coincide up to 0.2$nm$. This justifies that the choice of the method is not really critical, and that our results hold for both methods. The comparison of true experimental distributions with the distributions reconstructed from both methods is shown in figure SUP \ref{fit}.

\begin{table}[h!]
\begin{center}
\label{Table 1}
\begin{tabular}{| c |c | c |l c | c |}
\hline
\, & & & & & \\
 \, & $\mu_{mom} (nm)$ & $\sigma_{mom} (nm)$ & &  $\mu_{fit} (nm)$ & $\sigma_{fit} (nm)$  \\
\hline
$AAV8$ & 22.0 & 1.2 &  & 22.1 & 1.0 \\ 
\hline
$HBV$ & 24.1 & 2.1 &  & 24.3 & 2.1 \\  
\hline
\end{tabular}
\end{center}
\caption{\label{Table 1} Statistical moments determined either by the method of moments (first two columns) or by least-square fitting of the distribution}
\end{table}

\subsection{General result about the scaling model}
Our goal in this section is to derive the Eq. 9 of main text. This equation relates the second derivative of the energy with respect to height with the minimal energy and the value height at this minimal energy. In order to do so, we assume that the general form of the energy as function the indentation $\Delta=2R-H$ is 
\begin{equation}
E(\Delta)=A \Delta^{\alpha}-B \Delta^{\beta}
\end{equation}
where $\alpha$ and $\beta$ are two exponent that satisfy $\alpha>\beta$. This condition implies that there exist a minimum in the energy. The position of the minimum is obtained a in straightforward way as the point where the derivative of the energy vanishes. This point is
\begin{equation}
\Delta_{min}=\left(\frac{B\beta}{A\alpha}\right)^{\frac{1}{\alpha-\beta}}
\end{equation}
The minimal energy is obtained by replacing this value into the original energy. One gets
\begin{equation}
E_{min}=\frac{B^{\frac{\alpha}{\alpha-\beta}}}{A^{\frac{\beta}{\alpha-\beta}}}\left(\left(\frac{\beta}{\alpha}\right)^{\frac{\alpha}{\alpha-\beta}}-\left(\frac{\beta}{\alpha}\right)^{\frac{\beta}{\alpha-\beta}}\right)\sim\frac{B^{\frac{\alpha}{\alpha-\beta}}}{A^{\frac{\beta}{\alpha-\beta}}}
\end{equation}
The scaling approximation holds because the second term is a numerical prefactor, it does not have any physical dimension.
It is possible to rewrite the energy introducing the minimal energy and the minimal position:
\begin{equation}
E(\Delta)=\frac{E_{min}}{\left(\left(\frac{\beta}{\alpha}\right)^{\frac{\alpha}{\alpha-\beta}}-\left(\frac{\beta}{\alpha}\right)^{\frac{\beta}{\alpha-\beta}}\right)}\left(\left(\frac{\beta}{\alpha}\right)^{\frac{\alpha}{\alpha-\beta}}\left(\frac{\Delta}{\Delta_{min}}\right)^{\alpha}-\left(\frac{\beta}{\alpha}\right)^{\frac{\beta}{\alpha-\beta}}\left(\frac{\Delta}{\Delta_{min}}\right)^{\beta}\right)
\end{equation}
The estimation of the second derivative at the minimum gives, after some algebra:
\begin{equation}
\frac{\partial ^2 E}{\partial X^2}\bigg \|_{\Delta=\Delta_{min}}=\frac{E_{min}}{\Delta_{min}^2}\frac{\frac{\beta^{\frac{\alpha}{\alpha-\beta}}}{\alpha^{\frac{\beta}{\alpha-\beta}}}(\alpha-\beta)}{\left(\left(\frac{\beta}{\alpha}\right)^{\frac{\alpha}{\alpha-\beta}}-\left(\frac{\beta}{\alpha}\right)^{\frac{\beta}{\alpha-\beta}}\right)}\sim \frac{E_{min}}{\Delta_{min}^2}
\end{equation}
which demonstrates the scaling relation Eq.9 of the main text.

\subsection{Height dispersion and particle polydispersity}
In this section, we test whether it is possible to reproduce the height dispersion with particle polydispersity, or more quantitatively with capsid radius variation only. If we neglect the adhesion and deformation contribution, the numbers of height distribution of AAV8 and HBV implies that the relative variation of radius is 0.2 and 0.15 respectively. This change in radius can be converted in a change in the number of protein associated to radius increase for example. If we assume that the area of the equivalent sphere mimicking the viral capsid is proportional to the number of subunits, then the ratio of areas give the relative increase in the number of proteins in the process of radius variation.  In other words, if the radius changes from $R$ to $R(1+\alpha)$, the area changes from $4\pi R^2$ to $4\pi R^2(1+\alpha)^2$. The number of proteins changes from $\frac{4\pi R^2}{a^2}$ (where $a$ is the typical size of a protein in the capsid) to $\frac{4\pi R^2}{a^2}(1+\alpha)^2$. 
For AAV8, an increase of radius of $\alpha=0.2$ is associated with $(1+\alpha)^2-1\simeq 0.44$ variation of number of proteins. In others words, since AAV8 contains 60 proteins, more than 26 proteins would be needed in order to accommodate for such a radius variation. This number is obviously huge at the scale of a single capsid. It is not realistic with respect to the small polydispersity mentioned in the literature. Therefore, it seems safe to disregard the radius variation as influencing the height dispersion.

\bibliographystyle{apsrev4-2}  
\bibliography{virus}